\documentclass{raa}           
\usepackage{graphicx,times}
\usepackage{natbib}
\usepackage{amssymb,amsmath}
\bibpunct{(}{)}{;}{a}{}{,}

\usepackage[a4paper=true,dvipdfm=true,pagebackref=true]{hyperref}
\hypersetup{pdftitle = The title of my PDF, pdfauthor = My name, pdfsubject= The subject, pdfkeywords = keyword1 keyword2 keyword3} 
\hypersetup{colorlinks = true, linkcolor = green, anchorcolor = red, citecolor = blue, filecolor = red, pagecolor = red, urlcolor = red}

\begin{document}

   \title{Measurements of Ca II Infrared Triplet Lines
of Young Stellar Objects}

 \volnopage{ {\bf 2012} Vol.\ {\bf X} No. {\bf XX}, 000--000}
   \setcounter{page}{1}

   \author{Keiko Moto'oka\inst{1}, Yoichi Itoh\inst{2}}

   \institute{ The Saga Prefecture Space \& Science Museum,
16351, Nagashima, Takeo, Takeo, Saga 843-0021, Japan \\
        \and
             Nishi-Harima Astronomical Observatory, Center for Astronomy, 
             University of Hyogo, 
             407-2, Nishigaichi, Sayo, Hyogo 679-5313, Japan; 
             {\it yitoh@nhao.jp} \\
\vs \no
   {\small Received 2012 June 12; accepted 2012 July 27}
}

\abstract{
Equivalent widths and line widths of Ca II infrared triplet
emission lines were measured
in high-resolution optical spectra of 39 young stellar objects.
We found that the equivalent widths of the emission lines decrease with
stellar evolution.
It has been often claimed that strong chromospheric activity
is generated by a dynamo process caused by fast rotation of 
the photosphere.
However, we found no clear correlation between the strength of the 
Ca II lines
and the stellar rotation velocity.
Instead, we found that the objects with high mass accretion rates
had stronger Ca II emission lines.
This correlation supports 
the turbulent chromosphere model or 
the magnetic accretion theory for
classical T Tauri stars.
We also noticed that the equivalent widths of Ca II lines of 
transitional disk objects are one-tenth of those of classical T Tauri
stars, even if the masses of the circumstellar disks are comparable.
}

   \authorrunning{K. Moto'oka and Y. Itoh}            
   \titlerunning{Ca II IRT lines of YSOs}  
   \maketitle

%
\section{Introduction}           
\label{sect:intro}

The chromosphere is the region between the photosphere and the corona,
through which energy is transferred via regular and irregular
activities.
The chromosphere and its associated events are well investigated
for the nearest star, the Sun.
The Hinode satellite obtained space- and time-resolved images
of the solar chromosphere and revealed its energetic activity
(e.g., \citealt{Katsukawa2007}).

For main-sequence stars, chromospheric activity is often discussed 
in relation to the stellar rotation.
\cite{Noyes1984} found that objects with short rotation periods have
strong Ca II H \& K emission lines, which are an indicator of chromospheric 
activity.
They concluded that chromospheric activity is induced by the magnetic
field generated by the dynamo process.
The idea that chromospheres are magnetically driven is widely accepted for
solar-mass main-sequence stars,
and has also been applied to young stellar objects (YSOs). 
\cite{Neuhauser1995} found that
weak-line T Tauri stars (WTTSs) have stronger X-ray emission than
classical T Tauri stars (CTTSs).
Because WTTSs have a faster mean rotational velocity than CTTSs,
it is indicated that the strong dynamo process by fast rotation activates
the coronal region of the WTTSs.

Another indication of activity of YSOs was reported by
\cite{Hamann1992a}, who carried out an optical spectroscopy of 
53 T Tauri stars and 32 Herbig Ae/Be stars.
They detected narrow line emissions such as Ca II and Mg I.
They interpreted these data to mean that
the narrow lines are generated in stellar chromosphere.
In addition to the narrow line component, Ca II infrared triplet (IRT)
emission lines often have a broad line component.
This profile is well explained by the magnetospheric accretion model
(e.g. \citealt{Muzerolle1998}).

We investigated the activity of YSOs by
examining Ca II IRT emission.
The Ca II IRT lines are one of the strongest emission lines in an optical
spectrum of a low-mass YSO.
 They have only a small amount of interstellar extinction compared to 
Ca II H \& K lines in blue.

\section{Observations}

We carried out high-resolution optical spectroscopy of 12 T Tauri stars
using the High Dispersion Spectrograph (HDS) mounted on the Subaru Telescope.
The data were obtained on 2007 September 18 with the StdNIRb mode and
the 0\farcs6 width slit. 
These instrument settings achieved a wavelength coverage of 6650 -- 9360
\AA~ and a spectral resolution of $\sim$ 60,000.
The integration time for each object was between 600 s and 1500 s.

We also used archived data of 27 T Tauri stars that were
obtained with the High Resolution
Echelle Spectrometer (HIRES) mounted on the Keck Telescope.
The data were taken by S. E. Dahm on 2006 November 30, 2006 December 1,
2008 December 3, and 2008 December 4.
The wavelength coverage was between 4800 \AA~ and 9380 \AA~ and the 
spectral resolution was $\sim$ 70,000.
The integration time for each object was between 300 s and 1200 s.
We also used the Ca II IRT measurements of 6 T Tauri stars reported
in \cite{Hamann1992a}.
The objects we investigated are summarized in table \ref{target}.

\begin{table}
\begin{center}
\caption{Targets}
\label{target}
\begin{tabular}{lllll}
\hline
\hline
Object & Spectral Type$^{1}$ & Binary$^{2}$ & $R$-mag$^{3}$ & Telescope \\
\hline
\multicolumn{5}{c}{Classical T Tauri stars} \\
\hline
AA Tau & M0 & S & 11.80 & Keck/HIRES \\
BP Tau & K7 & S & 13.38 & Keck/HIRES \\
CW Tau & K5 & S & 11.75 & Keck/HIRES \\
CY Tau & M2 & S & 12.50 & Keck/HIRES \\
DF Tau & K5 & B & 13.34 & Keck/HIRES \\
DG Tau & G  & S & 11.40 & Keck/HIRES \\
DK Tau & M0 & B & 11.08 & Keck/HIRES \\
DL Tau & G  & S & 11.85 & Keck/HIRES \\
DO Tau & G  & S & 12.30 & Keck/HIRES \\
DR Tau & K4 & S & 10.68 & Keck/HIRES \\
FN Tau & M5 & S & 13.48 & Keck/HIRES \\
FP Tau & M2.5 & S & 12.13 & Keck/HIRES \\
FX Tau & M4 & B & 12.40 & Keck/HIRES \\
GI Tau & K5 & S & 12.15 & Keck/HIRES \\
GK Tau & K7 & B & 11.58 & Keck/HIRES \\
HL Tau & K9 & S & 10.63 & Hamann \& Persson \\
HN Tau & K5 & B & 13.31 & Keck/HIRES \\
LkCa 8 & M0 & S & 12.70 & Keck/HIRES \\
RW Aur & G5 & B & 10.06 & Hamann \& Persson \\
RY Tau & F8 & S & 10.20 & Hamann \& Persson \\
T Tau  & G5 & B & 9.19 & Hamann \& Persson \\
XZ Tau & G  & B & 13.56 & Hamann \& Persson \\
\hline
\end{tabular}
\end{center}
$^{1}$: SIMBAD database

$^{2}$: B: Binary, S: Single star.
\cite{Leinert1993}, \cite{Ghez1997},
\cite{Kohler1998},
\cite{Sartoretti1998},
\cite{Ireland2008},

$^{3}$: PPMXL catalog
\end{table}

\begin{table}
\addtocounter{table}{-1}
\begin{center}
\caption{(continued)}
\begin{tabular}{lllll}
\hline
\hline
Object & Spectral Type$^{1}$ & Binary$^{2}$ & $R$-mag$^{3}$ & Telescope \\
\hline
\multicolumn{5}{c}{Transitional disk objects} \\
\hline
CoKu Tau 4 & M1.5 & B & 12.10 & Keck/HIRES \\
CX Tau     & M1.5 & S & 12.63 & Keck/HIRES \\
DM Tau     & K5   & S & 13.61 & Keck/HIRES \\
FO Tau     & M2   & B & 14.00 & Keck/HIRES \\
GM Aur     & K5   & S & 11.02 & Keck/HIRES \\
LkCa 15    & K5   & S & 11.43 & Keck/HIRES \\
UX Tau     & G5   & B & 10.30 & Keck/HIRES \\
V773 Tau   & K2   & B &  7.49 & Keck/HIRES \\
V836 Tau   & K7   & S & 12.70 & Keck/HIRES \\
\hline
\multicolumn{5}{c}{Weak-line T Tauri stars} \\
\hline
HBC 374          & K7   & S & 11.21 & Subaru/HDS \\
HD 283716        & K0   & S & 9.79  & Subaru/HDS \\
NTTS 032641+2420 & K1   & S & 11.60 & Subaru/HDS \\
NTTS 041559+1716 & K7   & S & 11.50 & Subaru/HDS \\
NTTS 042417+1744 & K1   & S & 10.20 & Subaru/HDS \\
RX J0405.3+2009  & K1   & S & 9.57  & Subaru/HDS \\
RX J0409.2+1716  & M1   & S & 12.50 & Subaru/HDS \\
RX J0438.6+1546  & K2   & S & 10.27 & Subaru/HDS \\
RX J0452.5+1730  & K4   & S & 11.50 & Subaru/HDS \\
RX J0459.7+1430  & K4   & S & 10.80 & Subaru/HDS \\
V410 Tau         & K4   & B & 8.45  & Hamann \& Persson \\
V819 Tau         & K7   & S & 12.24 & Keck/HIRES \\
V827 Tau         & K7   & S & 11.39 & Subaru/HDS \\
V830 Tau         & M0-1 & S & 11.30 & Subaru/HDS \\
\hline
\hline
\end{tabular}
\end{center}
\end{table}

We reduced the HDS data using the following standard steps: 
overscan subtraction,
bias subtraction, flat fielding, removal of scattered light, extraction
of a spectrum, wavelength calibration using a Th-Ar lamp,
and continuum normalization.
We used IRAF packages for all data-processing procedures.
A detail description of the data-reduction method 
is presented in \cite{Takagi2011}.
The HIRES data were reduced with the Mauna Kea Echelle Extraction (MAKEE)
package.

Emission lines of the Ca II IRT 
are superimposed on broad photospheric absorptions
of Ca II.
To construct spectra in which the equivalent widths of these emission lines are 
measured,
we filled the absorption features using a spectrum of a dwarf with
the same spectral type.
We obtained the dwarf spectra from the HIRES data archive.
Because the majority of YSOs are fast rotators,
the dwarf spectra were convolved with a Gaussian profile so that the
FWHMs of the photospheric absorption lines were comparable to those of the
YSO spectra.
Each convolved dwarf spectrum was subtracted from the corresponding
YSO spectrum, then
add unity.
This procedure mostly removed the broad Ca II absorptions 
as well as the other photospheric
absorption lines.
Although it is possible that veiling effects caused by circumstellar materials 
could also alter the YSO emission spectra,
we did not compensate for this because the veiling effect is insignificant
in the $I$-band (\citealt{Bertout1988}).

We measured the equivalent widths, line widths,
and radial velocities
of Ca II IRT emission lines by fitting the line profiles with a Voigt 
function.
Before subtracting the dwarf spectra, 
we obtained the rotational velocity ($v\sin i$) of each YSO
by measuring the FWHM of an unblended photospheric absorption line
of Ti at 8683 \AA.
For objects whose values were taken from \cite{Hamann1992a},
the radial velocities of the Ca II IRT emissions and the photospheric
rotational velocity are unknown.

\section{Results}

\begin{figure}
 \centering
 \includegraphics[width=14.0cm, angle=0]{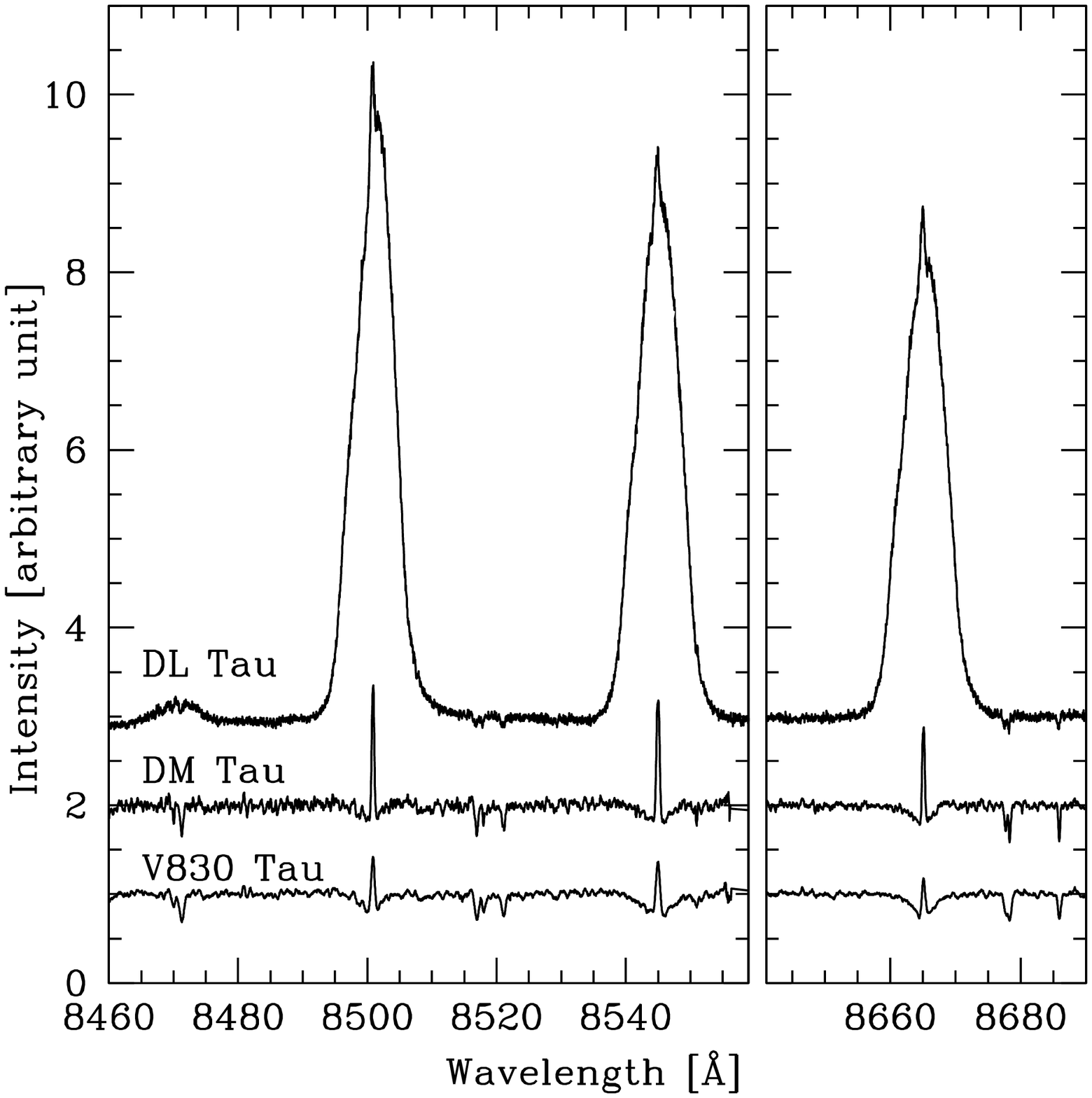}
 \caption{Ca II IRT emissions of YSOs. DL Tau, a CTTS, has
prominent emission features. DM Tau, a transitional disk object, and V830 Tau, a
 WTTS,
show narrow and weak emission lines superimposed on broad absorption features.
Photospheric absorption lines are not subtracted.}
 \label{spec}
\end{figure}

Figure \ref{spec} shows example spectra from a CTTS,
a transitional disk object, and a WTTS.
In these examples, photospheric absorption lines are not subtracted.
The spectrum of the CTTS shows the strongest and broadest emission lines
for the Ca II IRT, completely filling the absorption components.
The transitional disk object has narrow emission lines,
and the WTTS shows weak emission lines.
Table \ref{ew} lists the equivalent widths, the FWHMs, and the radial velocities
computed from the emission lines as well as the rotational velocity of the 
photosphere, the mass accretion rate, and the mass of the circumstellar
disk.

\begin{table}
\begin{center}
\caption{Measurements of Ca II infrared triplet lines}
\label{ew}
\tiny
\begin{tabular}{p{70pt}cccccccccccc}
\hline
\hline
Object & 
\multicolumn{3}{c}{Equivalent width} &
\multicolumn{3}{c}{Line width} &
\multicolumn{3}{c}{Radial velocity} &
$v_{\rm rot}$ &
log \. M  &
log M$_{\rm disk}$  \\
&
\multicolumn{3}{c}{[\AA]} &
\multicolumn{3}{c}{[km s$^{-1}$]} &
\multicolumn{3}{c}{[km s$^{-1}$]} &
[km s$^{-1}$] &
{\tiny [M$_{\odot}$ yr$^{-1}$]} &
[M$_{\odot}$] \\
& 8498 \AA & 8542 \AA & 8662 \AA &
 8498 \AA & 8542 \AA & 8662 \AA &
 8498 \AA & 8542 \AA & 8662 \AA \\
\hline
\multicolumn{13}{c}{Classical T Tauri stars} \\
\hline
AA Tau &  1.47 &  4.25 &  3.12 &  25 &  33 &  28 &   -1 &   1 &   2 &  16 & -8.48 & -2.00 \\
BP Tau &  6.37 &  8.67 &  6.44 &  27 &  39 &  30 &    1 &   2 &   3 &  15 & -7.54 & -1.70 \\
CW Tau & 11.27 & 11.13 &  9.31 & 184 & 219 & 190 &   11 &  20 &  16 &  38 & -7.61 & -2.70 \\
CY Tau &  6.45 &  8.39 &  6.43 &  34 &  65 &  53 &   -2 &  -1 &   0 &  14 & -8.12 & -2.22 \\
DF Tau &  2.94 &  3.15 &  2.40 &  61 &  66 &  56 &   -1 &  -1 &   0 &  33 & -7.62 & -3.40 \\
DG Tau & 38.77 & 38.04 & 35.17 & 196 & 213 & 191 &    4 &   4 &   4 &  41 & -6.30 & -1.70 \\
DK Tau &  6.85 &  6.80 &  6.44 & 140 & 166 & 142 &   -2 & -10 &   1 &  18 & -7.42 & -2.30 \\
DL Tau & 53.52 & 52.37 & 43.04 & 234 & 268 & 241 &    6 &   6 &   4 &  15 & -6.79 & -1.05 \\
DO Tau & 28.02 & 30.16 & 19.56 & 126 & 137 & 119 &   -3 &  -1 &   0 &  18 & -6.85 & -2.15 \\
DR Tau & 25.94 & 27.13 & 24.15 & 108 & 158 & 126 &    1 &  -6 &   3 & 116 & -6.50 & -1.70 \\
FN Tau &  1.17 &  1.95 &  1.25 &  17 &  26 &  19 &    0 &   1 &   2 &  10 \\
FP Tau &  0.66 &  1.11 &  0.97 &  54 &  58 &  51 &  -21 & -20 & -18 &  46 \\
FX Tau &  0.67 &  1.73 &  1.25 &  23 &  29 &  21 &    0 &   0 &   3 &  13 & -8.65 & -3.05 \\
GI Tau &  1.19 &  2.30 &  1.68 &  22 &  27 &  22 &   -1 &   0 &   2 &  16 & -8.02 \\
GK Tau &  1.70 &  2.16 &  1.85 &  39 &  44 &  36 &    2 &   3 &   5 &  27 & -8.19 & -2.70 \\
HL Tau & 31.70 & 31.80 & 25.20 & 148 & 175 & 163 &      &     &     &     &     & -1.22 \\
HN Tau & 50.96 & 46.66 & 39.21 & 321 & 309 & 276 & -100 &  58 &  64 &     & -8.89 & -3.10 \\
LkCa 8 &  4.76 &  5.41 &  3.14 &  28 &  34 &  27 &    2 &   4 &   5 &  16 & -9.10 & -2.52 \\
RW Aur & 69.10 & 68.00 & 53.20 & 245 & 310 & 291 &      &     &     &     & -7.12 & -2.40 \\
RY Tau &  3.76 &  2.64 &  2.57 & 250 & 325 & 320 &      &     &     &     & -7.11 & -1.70 \\
T Tau  &  4.19 &  4.34 &  3.87 &  80 &  83 &  82 &      &     &     &     & -7.12 & -2.10 \\
XZ Tau &  8.43 &  8.09 &  6.53 &  91 &  94 &  99 &      &     &  \\
\hline
\multicolumn{13}{c}{Transitional disk objects} \\
\hline
CoKu Tau 4 & 0.46    &  0.65 & 0.48 & 44 & 50 & 38 &  11 &  10 &  15 & 37 & $<$-10.00 & -3.30 \\
CX Tau     & 1.29    &  1.89 & 1.20 & 34 & 40 & 32 &   4 &   5 &   6 & 29 & -8.97     & -3.00 \\
DM Tau     & 1.23    &  1.58 & 1.12 & 18 & 25 & 19 &  -1 &   0 &   1 & 11 & -7.95     & -1.70 \\
FO Tau     & 2.03    &  2.94 & 2.35 & 25 & 37 & 28 &   0 &   1 &   2 & 14 & -7.90     & -3.22 \\
GM Aur     & 1.50    &  3.65 & 2.91 & 30 & 43 & 41 &   0 &   1 &   2 & 20 & -8.02     & -1.52 \\
LkCa 15    & 0.91    &  1.23 & 0.96 & 25 & 30 & 11 &   2 &   4 &   6 & 18 & -8.87     & -1.30 \\
UX Tau     & 0.63    &  1.06 & 0.89 & 41 & 46 & 37 &  -2 &  -1 &   1 & 35 & -9.00     & -2.30 \\
V773 Tau   & $<$0.94 &  1.68 & 1.36 & 67 & 92 & 71 & -37 & -19 & -30 & 58 & -9.62     & -3.30 \\
V836 Tau   & 1.70    &  5.20 & 3.06 & 31 & 41 & 32 &   0 &   0 &   2 & 18 & -8.98     & -2.00 \\
\hline
\multicolumn{13}{c}{Weak-line T Tauri stars} \\
\hline
HBC 374          & 1.54 & 2.11 & 0.70 & 32  &  37 &  31 &  -1 &   1 &   1 &  24 & $<$-7.78 & $<$-3.40 \\
HD 283716        &      &      &      &     &     &     &  74 &  73 &  72 \\
NTTS 032461+2420 & 0.46 & 1.20 & 0.64 & 89  & 140 &  22 & -32 & -31 & -53 \\
NTTS 041559+1716 & 1.83 & 4.36 & 1.58 & 133 & 135 & 101 &   9 &   4 & -17 & 107 & $<$-8.92 & $<$-3.49 \\
NTTS 042417+1744 & 0.29 & 0.43 & 0.36 & 24  &  26 &  21 &   8 &   4 &   7 &  23 & $<$-8.03 & $<$-3.52 \\
RX J0405.3+2009  & 0.37 & 0.51 & 0.36 & 34  &  34 &  31 &   5 &   2 &   5 &  39 \\
RX J0409.2+1716  & 1.02 & 1.68 & 1.00 & 99  & 104 & 100 &  19 &  36 &  10 \\
RX J0438.6+1546  & 0.57 & 0.88 & 0.71 & 38  &  41 &  34 &   6 &   4 &   2 &  38 \\
RX J0452.5+1730  & 0.59 & 0.73 & 0.49 & 20  &  23 &  19 &   3 &   1 &   1 &  14 \\
RX J0459.7+1430  & 0.44 & 0.75 & 0.43 & 23  &  26 &  22 &   3 &   5 &   3 &  22 \\
V410 Tau         & 0.21 & 0.43 & 0.27 & 51  &  54 &  52 &     &     &     &     & $<$-8.42 & $<$-3.40 \\
V819 Tau         & 1.44 & 2.73 & 2.44 & 25  &  36 &  28 &   0 &   2 &   3 &  14 & $<$-8.48 & $<$-3.40 \\
V827 Tau         & 1.76 & 2.12 & 1.83 & 37  &  41 &  34 &   1 &   2 &   4 &  28 & $<$-8.15 & $<$-3.52 \\
V830 Tau         & 0.56 & 0.88 & 0.58 & 52  &  52 &  47 & -21 & -15 & -17 &  43 & $<$-8.10 & $<$-3.52 \\
\hline
\end{tabular}
\end{center}
\end{table}

\section{Discussion}

\begin{figure}
 \centering
 \includegraphics[width=14.0cm, angle=0]{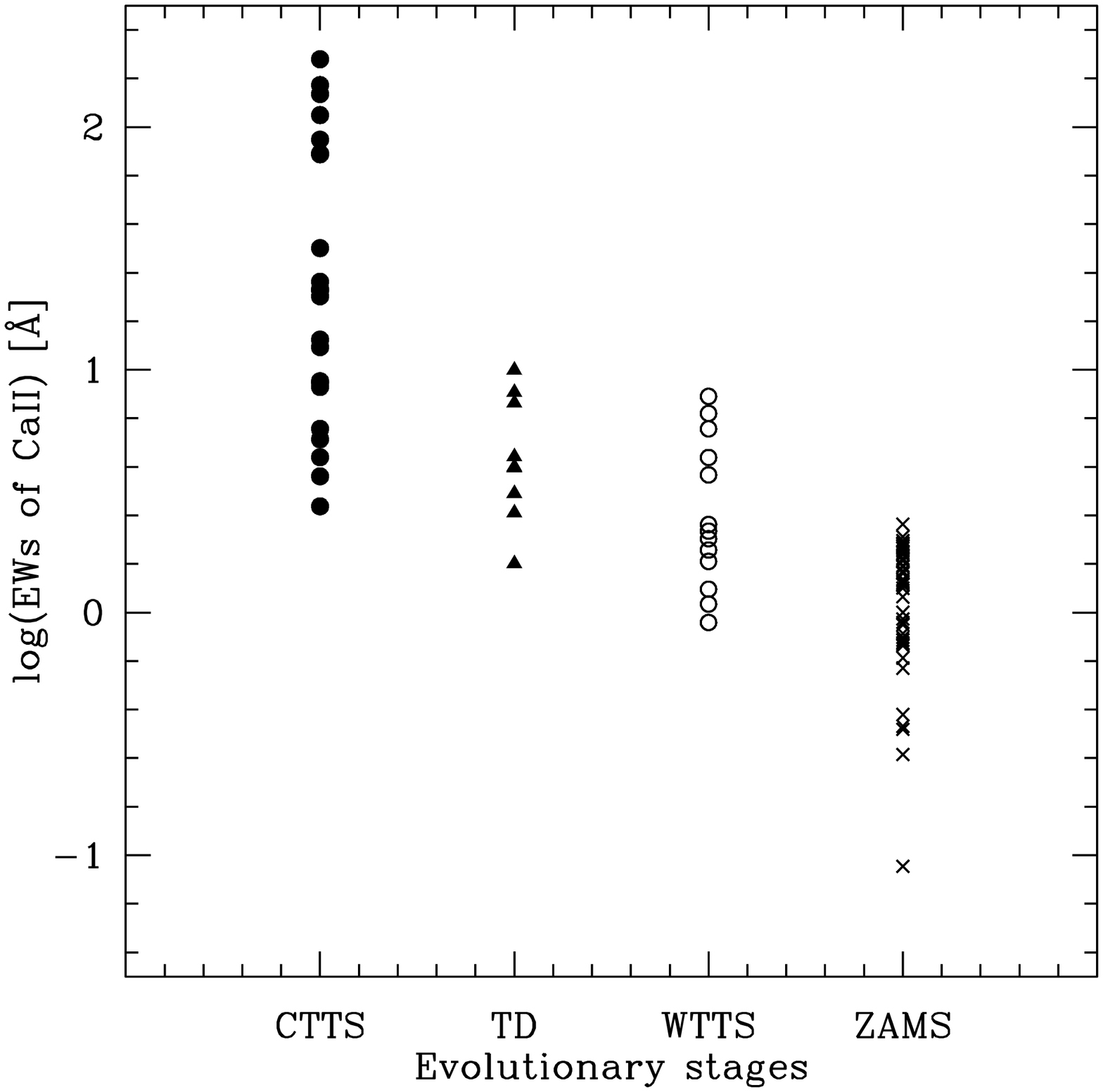}
 \caption{Equivalent widths of the Ca II IRT emission lines
as a function of the evolutional sequence. 
The vertical axis represents the sum of the equivalent widths
of the 3 emission lines.
As seen in Figure 1, CTTSs show
large equivalent widths.
}
 \label{EvEW}
\end{figure}

\begin{figure}
 \centering
 \includegraphics[width=14.0cm, angle=0]{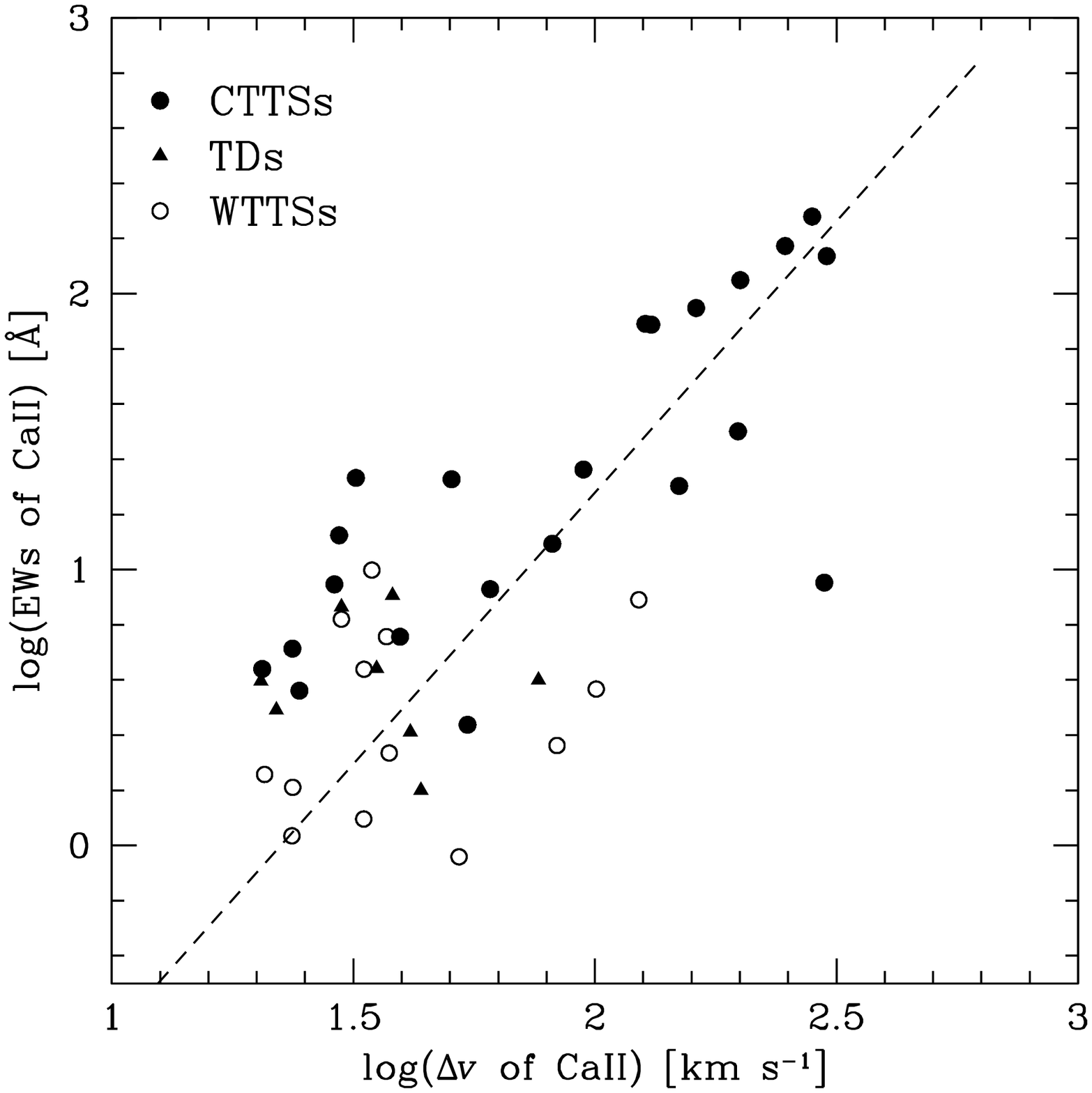}
 \caption{The equivalent widths of the Ca II IRT emissions as 
a function of the line widths of the emission lines.
The vertical axis represents the sum of the equivalent widths
of the 3 emission lines.
The horizontal axis means the average of the FWHMs of the 3 emission lines.
The dashed line indicates that the equivalent widths increase as the 
square of the line widths.
}
 \label{FWHMEW}
\end{figure}

Our results indicate that the Ca II IRT emission lines weaken with 
increasing stellar age.
Figure \ref{EvEW} shows the equivalent widths of the Ca II IRT emission lines
as a function of the evolutional sequence of the objects.
As a reference, we also show the equivalent widths of the lines
of ZAMSs reported by \cite{Marsden2009}, 
in which 24 solar-type stars in the IC 2391 cluster 
and 28 stars in the IC 2602 cluster are investigated.
Among the sample, CTTSs show the strongest emissions, although there
is a large scatter in the equivalent widths.
The emission lines of the WTTSs and ZAMSs
are weak.
A similar trend was reported in \cite{Hamann1992b},
in which T Tauri stars with large infrared 
excess show strong Ca II IRT emission.
Figure \ref{FWHMEW} shows the correlation between the FWHMs of the 
Ca II IRT lines and their equivalent widths.
Lines with stronger emissions are broader.
\cite{Batalha1996} proposed that the narrow line emits from the
hot chromosphere, while \cite{Hartmann1994} claimed that
the broad line is formed at magnetosphere in which 
circumstellar material falls onto
the central star.
A similar but weaker correlation was also found in the H$\alpha$ emission line
(\citealt{Reipurth1996}).

\begin{figure}
 \centering
 \includegraphics[width=14.0cm, angle=0]{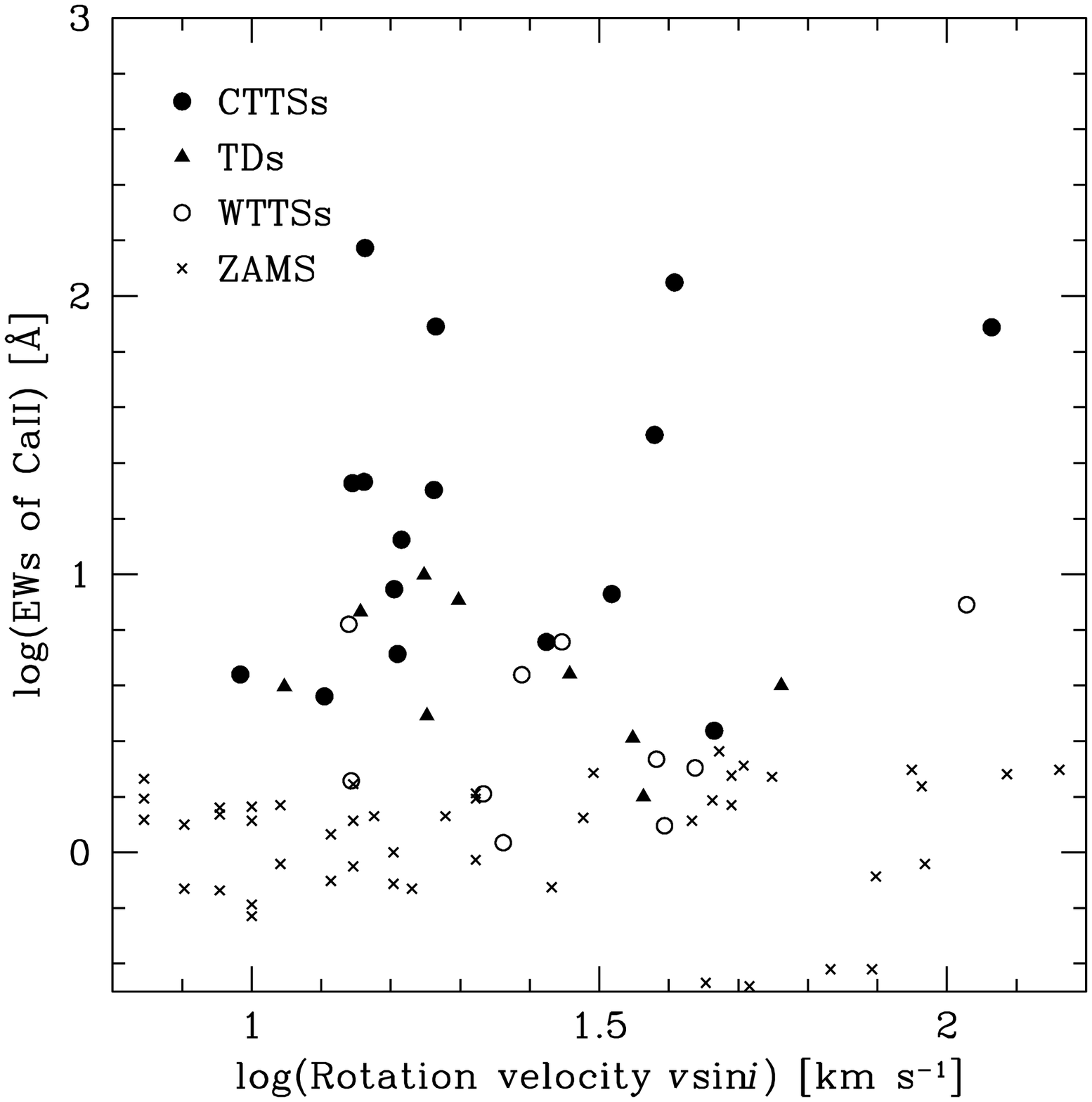}
 \caption{The equivalent widths of the Ca II IRT emissions 
of YSOs as a function of photospheric rotation velocity.
The vertical axis represents the sum of the equivalent widths
of the 3 emission lines.
There is no apparent correlation in these parameters.
}
 \label{rotEW}
\end{figure}

It is widely accepted that a strong dynamo process is caused
by a rapidly rotating photosphere.
Figure \ref{rotEW} shows the relationship between
the rotational velocities ($v\sin i$) and
the equivalent widths of the Ca II IRT emission lines. 
We see no correlation in these parameters.
\cite{Bouvier1990} and \cite{Neuhauser1995} pointed out
that WTTSs rotate faster than CTTSs.
Therefore, if chromospheric activity is induced
by the dynamo process, 
and if the Ca II IRT emission lines are of chromospheric origin,
then the Ca II IRT emission lines of WTTSs
are expected to be stronger than those of CTTSs.
Because the Ca II IRT emission lines of the WTTSs 
are not stronger than those of the CTTSs,
it does not appear that the strong Ca II emission lines
observed in CTTSs
are mainly generated in active chromosphere
induced by the dynamo process.

\begin{figure}
 \centering
 \includegraphics[width=14.0cm, angle=0]{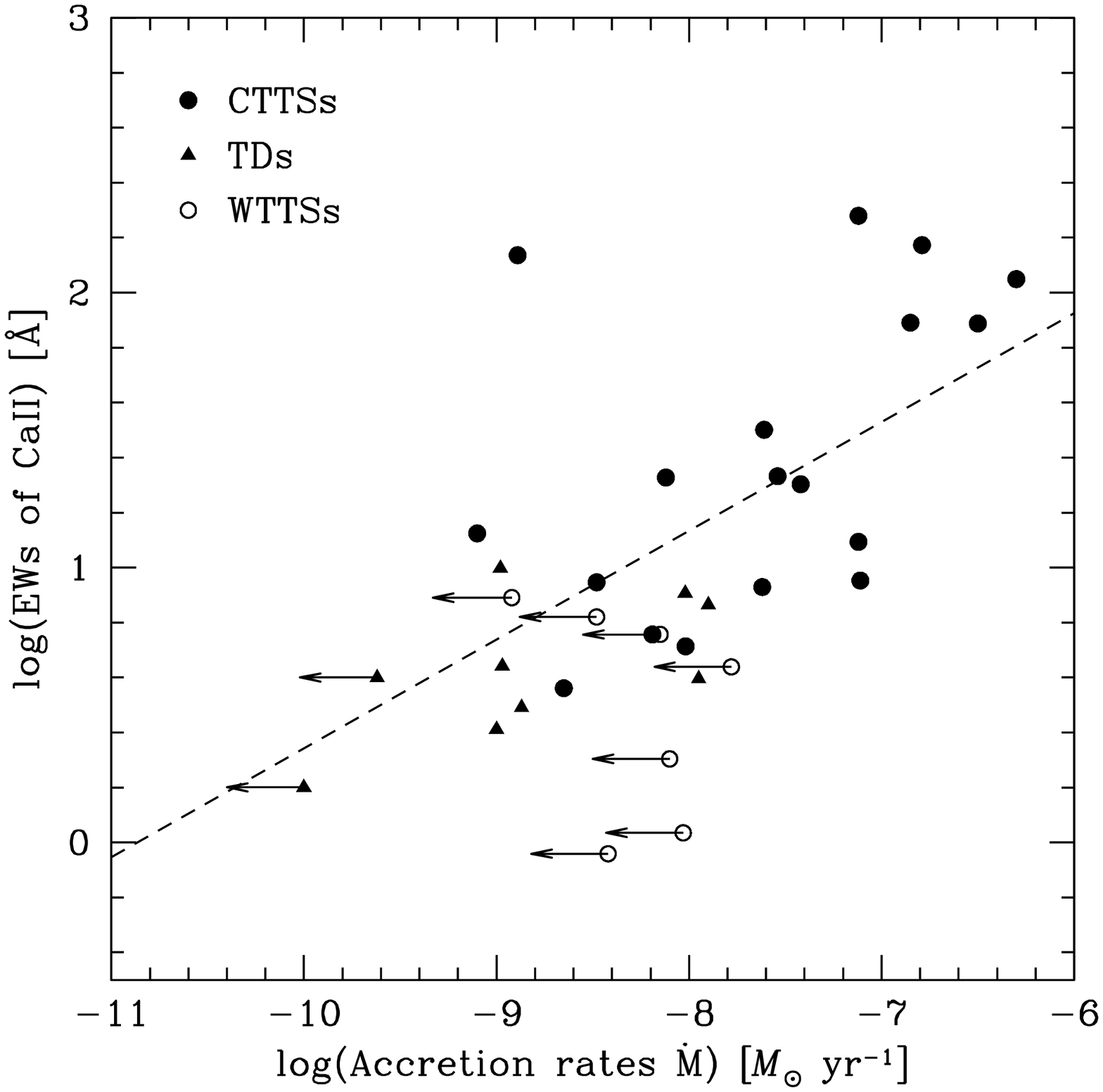}
 \caption{The equivalent widths of the Ca II IRT emission lines
of YSOs as a function of mass accretion rate of the circumstellar materials.
The vertical axis represents the sum of the equivalent widths
of the 3 emission lines.
The dashed line means that the equivalent widths increase as the mass accretion
rates to the power of 0.4.
}
 \label{AccEW}
\end{figure}

It is also well known that CTTSs usually have
high rates of mass accretion
from their circumstellar disks.
Figure \ref{AccEW} shows the relationship between
mass accretion rate and the equivalent widths of the Ca II IRT emissions.
We used the mass accretion rates from \cite{Gullbring1998},
\cite{Hartmann1998}, and \cite{White2001}. 
\cite{Gullbring1998} conducted intermediate-resolution
spectrophotometry from 3200 to 5200 \AA~ for 26 CTTSs and 3 WTTSs.
The veiling effect of a hot continuum caused by mass accretion phenomena
is not negligible in the $U$- and $B$-bands.
The amount of the veiling was estimated
by measuring the line-to-continuum ratio of a number of absorption
features of the YSO spectrum and the ratios of a template dwarf spectrum.
The line-to-continuum ratio was the flux ratio of an absorption
line and its adjacent continuum.
Then, they estimated the temperature of the hot continuum
region by comparing the fluxes at 3600 \AA, 4000 \AA, and
4750 \AA.
They assumed a constant temperature.
The extinction of the object was estimated from broad-band photometry, 
then they derived the luminosity of the hot continuum region, i.e., 
an accretion luminosity.
Finally, assuming a disk inner radius of 5 stellar radius,
they estimated the mass accretion rates.
They also proposed the relationship between the $U$-band excess
and the accretion luminosity determined from optical 
veiling estimate.
\cite{Hartmann1998} and \cite{White2001} calculated the mass accretion 
rates using this relationship.
From figure \ref{AccEW} we noticed a correlation between the mass 
accretion rates and the equivalent widths of the Ca II IRT emission lines;
objects with high mass accretion rates have strong emissions of the Ca II
IRT.
Models of magnetospheric accretion are proposed to explain broad
components of the emission lines observed in CTTSs.
In the model, the magnetic field connecting between the photosphere
and the circumstellar disk is expected (e.g. \citealt{Uchida1985}).

\begin{figure}
 \centering
 \includegraphics[width=14.0cm, angle=0]{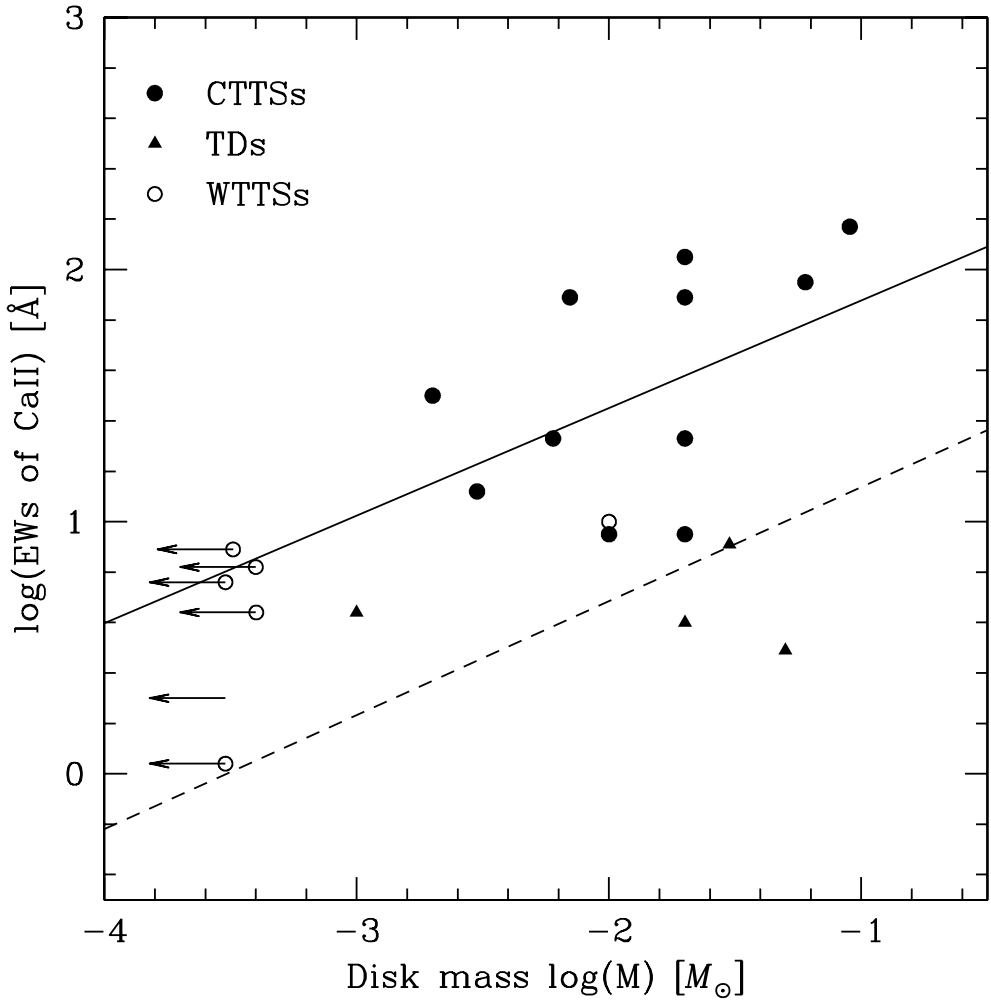}
 \caption{The equivalent widths of the Ca II IRT emission lines
of YSOs as a function of mass of the circumstellar disks. 
The vertical axis represents the sum of the equivalent widths
of the 3 emission lines.
Least-squares fits are plotted for the CTTS sample (solid line) 
and the transitional disk object sample (dashed line).
Both samples consist of single stars.
}
 \label{diskEW}
\end{figure}

\begin{figure}
 \centering
 \includegraphics[width=14.0cm, angle=0]{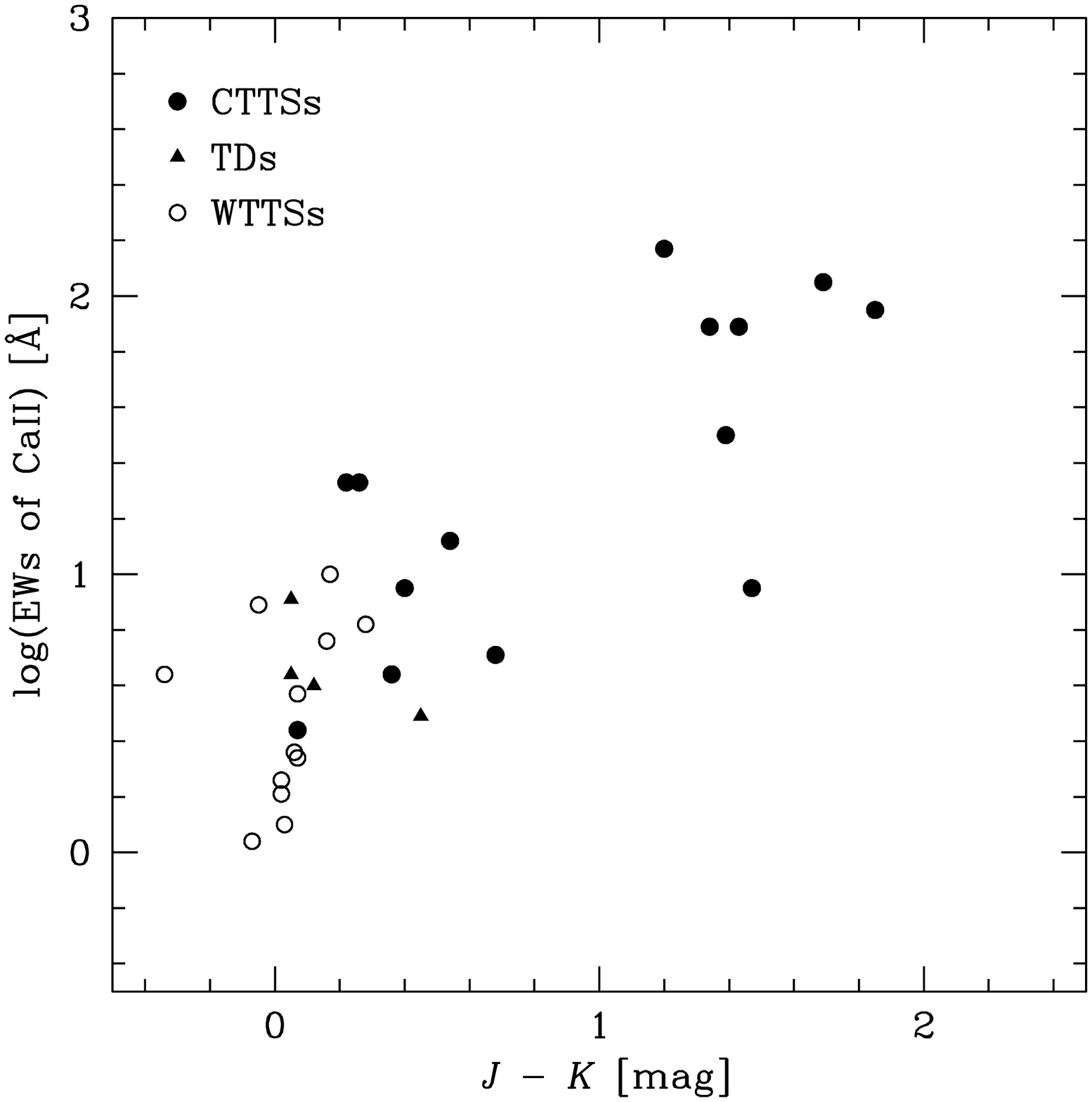}
 \caption{The equivalent widths of the Ca II IRT emission lines
of YSOs as a function of near-infrared excess. 
The vertical axis represents the sum of the equivalent widths
of the 3 emission lines.
The sample consists of single stars.
}
 \label{nirEW}
\end{figure}

The material is accreted from the circumstellar disk along the magnetic
field lines.
We investigated the correlation between the strengths of the Ca II IRT 
emission lines and the
mass of the circumstellar disk (Figure \ref{diskEW}).
We used the disk mass from \cite{Andrews2005}.
They conducted an extensive submillimeter continuum survey
of YSOs in the Taurus-Auriga star forming region.
For the objects whose mid- and far-infrared photometric data were available,
they fitted the spectrum of the model disk between mid-infrared and
submillimeter wavelengths,
then determined the disk mass.
For the other objects, the disk mass was estimated from the submillimeter 
flux.
We found that the equivalent widths of the Ca II IRT emission lines
of the transitional disk
objects are one-tenth the value of those of CTTSs, even if the masses of the
circumstellar disks are comparable.
It is considered that the transitional disk objects have an inner hole with a
radius of a few AU or a few tens of AU.
An object exhibits infrared excess if its disk fills such an inner region.
Figure \ref{nirEW} shows the equivalent widths of the Ca II IRT emission
lines as a function of near-infrared $J-K$ color.
Photospheric color of the object and reddening caused by interstellar 
material are subtracted, so that the color in the figure indicates
near-infrared excess caused by an inner region of the circumstellar disk.
The CTTSs show large variety in the near-infrared excess, whereas 
the transitional disk objects and the WTTSs have little near-infrared
excesses.
The objects with large near-infrared excess show strong Ca II IRT emissions.
This correlation supports an idea that materials accreting from the warmer
inner disk enhance the stellar activity.
We also suggest that the magnetic fields between the photosphere and
the circumstellar disk are already disconnected at the evolution
phase of the transitional disk objects.

\begin{figure}
 \centering
 \includegraphics[width=14.0cm, angle=0]{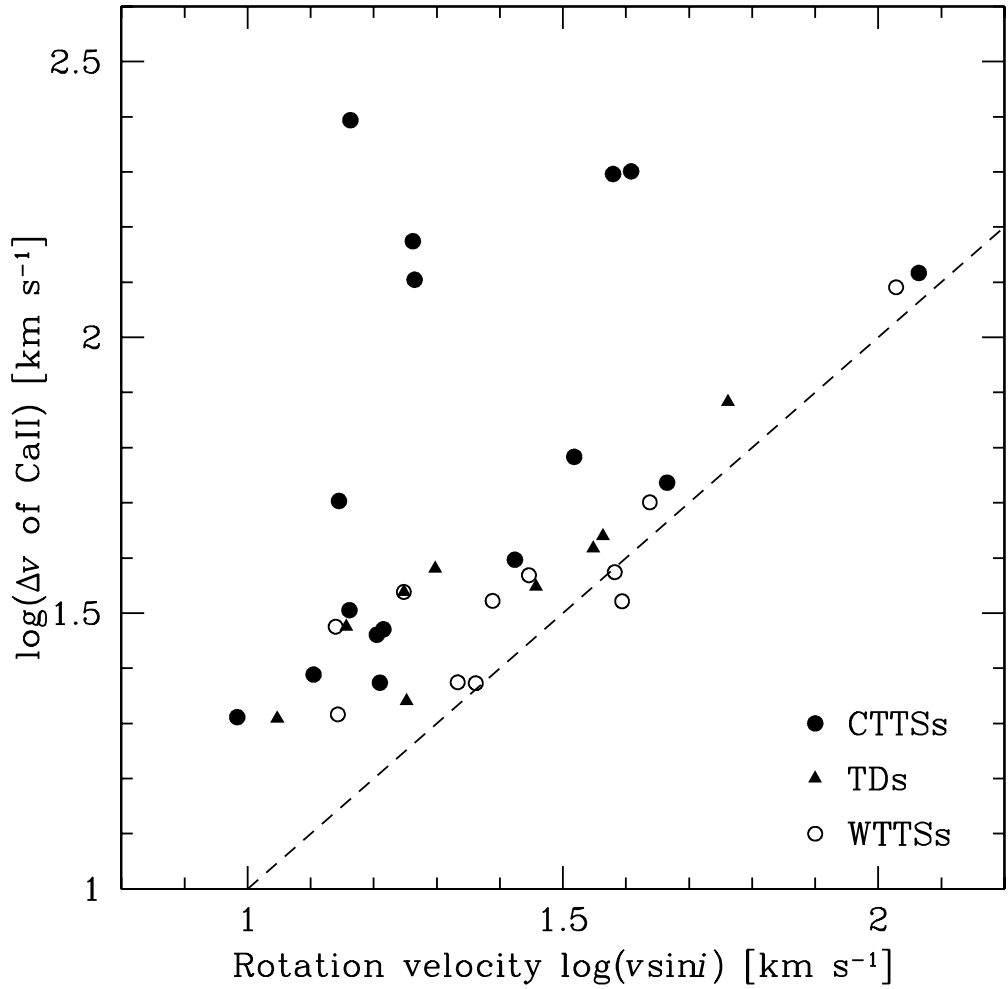}
 \caption{FWHM of the Ca II IRT emissions as a function of the FWHMs of
the photospheric absorption line.
The dashed line represents the case where the FWHMs of both lines are same.
}
 \label{rotFWHM}
\end{figure}

The line widths of the Ca II IRT emissions also differ significantly
between the CTTSs and the transitional disk objects.
Figure \ref{rotFWHM} shows the line widths of the Ca II IRT emissions as a
function of the rotational velocity of the photosphere.
The rotational velocity is
estimated from the line width of the absorption line.
As shown in the figure, the objects can be classified into two groups.
One group consists of the objects for which
the widths of the Ca II IRT emission lines are comparable
to, or slightly larger than, the width of the absorption line.
The most CTTSs as well as
all transitional disk objects and WTTSs are classified into this group.
We consider that the emission lines emanate from the stellar chromosphere
for the objects in this group.
The other group consists of the exceptions. All objects in this group are
CTTSs (CW Tau, DG Tau, DK Tau, DL Tau, and DO Tau).
The widths of the emission lines
are significantly broad, compared to the line widths
of the photospheric absorption line.
We found that the objects in this group 
have large mass accretion rates (\.M$>10^{-8}$
M$_{\odot}$ yr$^{-1}$, see table \ref{ew}).
We consider that the Ca II IRT emission lines of the objects in this group
emanate from the
magnetosphere between the photosphere and the circumstellar disk.
\cite{Muzerolle1998} observed 11 CTTSs with a spectral resolution 
$R \sim 35,000$.
Several emission lines including the Ca II IRT emission lines
were detected.
They classified the objects into three groups based on the shapes
of the Ca II IRT emission lines.
One is the objects with a narrow emission line superimposed on
the photospheric broad absorption.
The second group consists of the objects showing
both broad and narrow component emissions.
The objects belong to the third group show only a broad-component emission.
\cite{Hamann1992a} proposed that the
broad components of the Ca II IRT emission emanates from an extended envelope
with large turbulent velocities.
\cite{Muzerolle1998} constructed the magnetospheric model.
The magnetospheric emission line is characterized
by a large line width, a blueshifted asymmetry, and a slightly blueshifted
peak.
They claimed that the Ca II IRT emission line profiles of a part of the objects
in the third group are well reproduced by the magnetospheric model.
Among the exceptions in our sample (CW Tau, DG Tau, DK Tau, DL Tau, and
DO Tau), DG Tau and DO Tau exhibit the emission line profiles
similar to that given by the magnetospheric model.
However, the line profiles of the rest objects are complicated and
seem not to be reproduced by the magnetospheric model.
Moreover, by comparing our spectra to the spectra presented in
\cite{Muzerolle1998}, we find significant difference in the line profiles
between the spectra.
This difference may indicate variability of the emission lines.
Because the spatial sizes of the chromosphere and the 
magnetosphere are different,
time-series spectroscopy with short intervals will identify whether
the origin of the broad and strong emission lines of the
Ca II IRT is the turbulent chromosphere or the magnetosphere.

\section{Conclusions}

We measured strengths and line profiles of Ca II infrared triplet
emission lines with high-resolution optical spectra of 39 young stellar
objects.
\begin{enumerate}
\item The equivalent widths of the emission lines decrease with
stellar evolution.
\item The CTTSs with high mass accretion rates
show strong and broad emission of the Ca II lines.
It is considered that the lines emit from 
the turbulent chromosphere or
the magnetosphere between the
photosphere and the circumstellar disk.
\item  The transitional disk objects and the WTTSs show weak and narrow
emission of the Ca II lines. The line widths are comparable to, or
slightly larger than, the line width of the photospheric absorption line.
The emission lines are attributed to the chromospheric activity.
The equivalent widths of Ca II lines of the 
transitional disk objects are one-tenth of those of classical T Tauri
stars, even if the masses of the circumstellar disks are comparable.
Mass accretion from the warmer inner disk induces chromospheric
and/or magnetospheric activities.
\end{enumerate}

\begin{acknowledgements}
We thank the telescope staff members and operators at the Subaru Telescope.
This research has made use of the Keck Observatory Archive (KOA), 
which is operated by the W. M. Keck Observatory and the NASA Exoplanet 
Science Institute (NExScI), under contract with the National Aeronautics 
and Space Administration. 
\end{acknowledgements}

\clearpage
\appendix

\section{Spectra of the YSOs}

High resolution spectra of 39 YSOs are presented in the appendix.

\begin{figure}[h]
 \centering
 \includegraphics[width=14.0cm, angle=0]{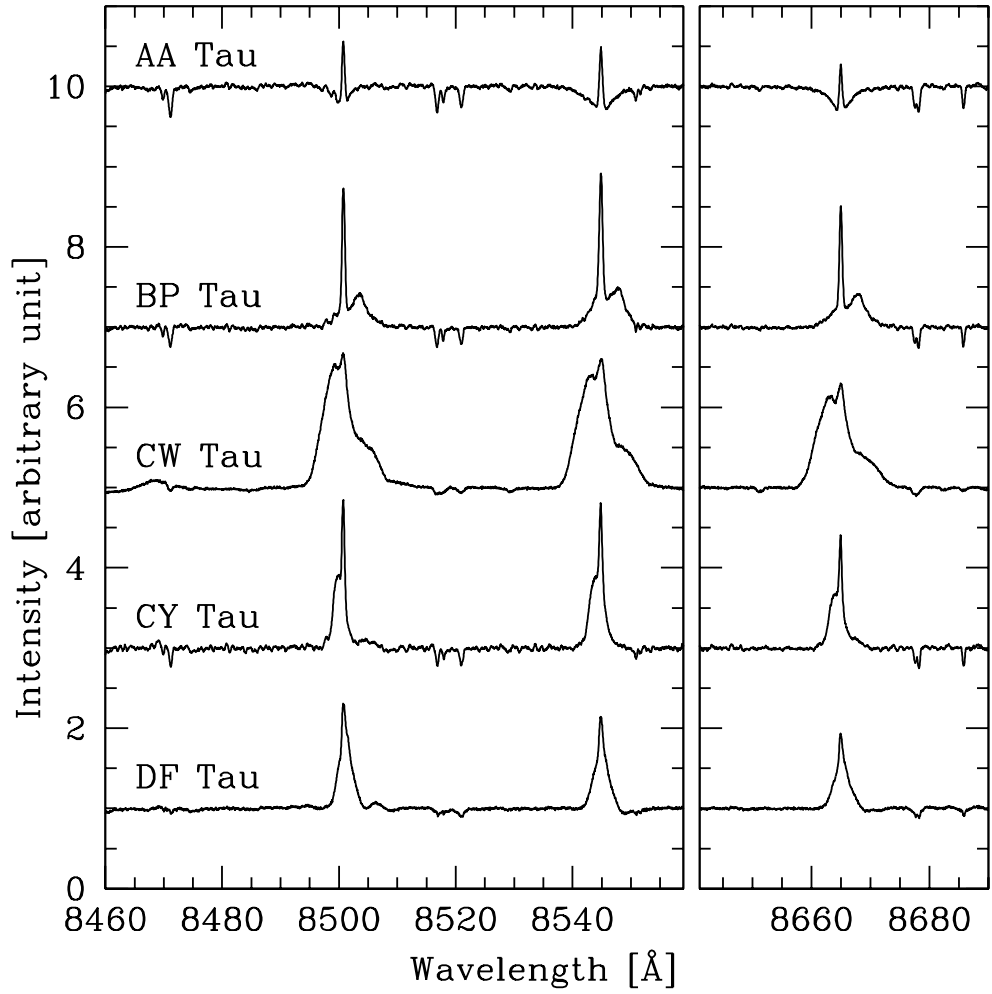}
 \caption{High resolution spectra of the YSOs.}
 \label{ap1}
\end{figure}

\begin{figure}[h]
 \centering
 \includegraphics[width=14.0cm, angle=0]{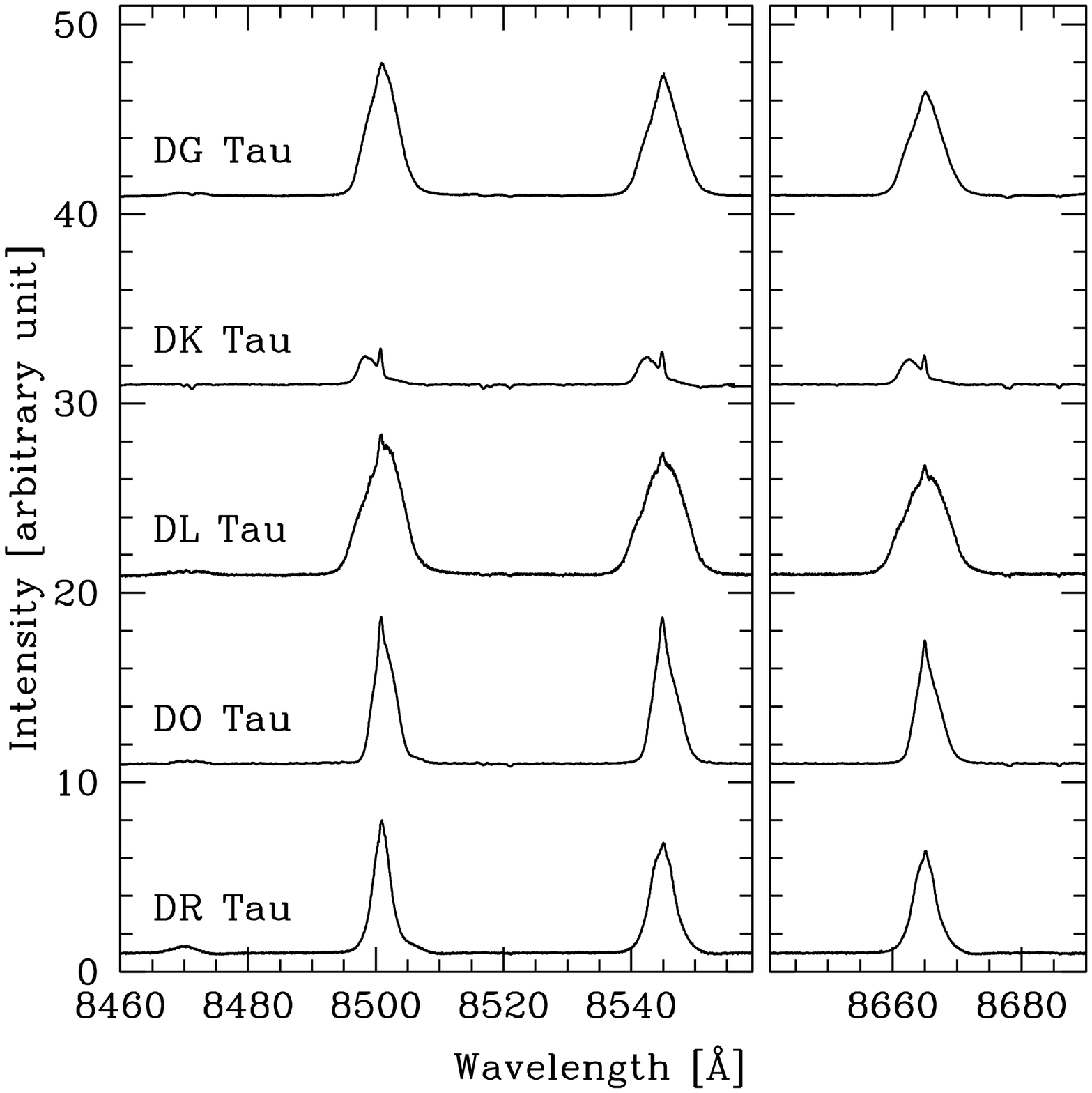}
 \caption{High resolution spectra of the YSOs.}
 \label{ap2}
\end{figure}

\begin{figure}[h]
 \centering
 \includegraphics[width=14.0cm, angle=0]{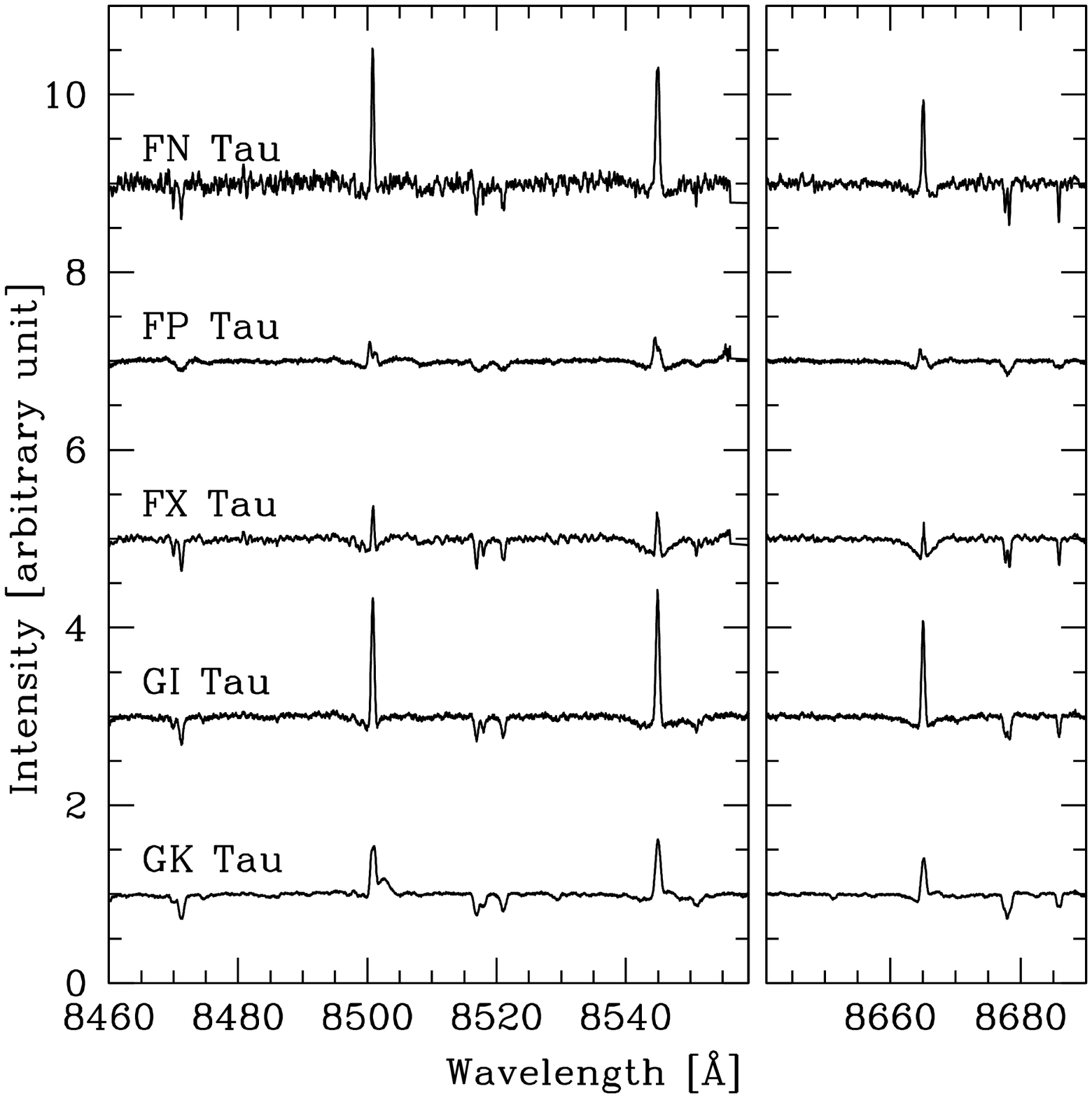}
 \caption{High resolution spectra of the YSOs.}
 \label{ap3}
\end{figure}

\begin{figure}[h]
 \centering
 \includegraphics[width=14.0cm, angle=0]{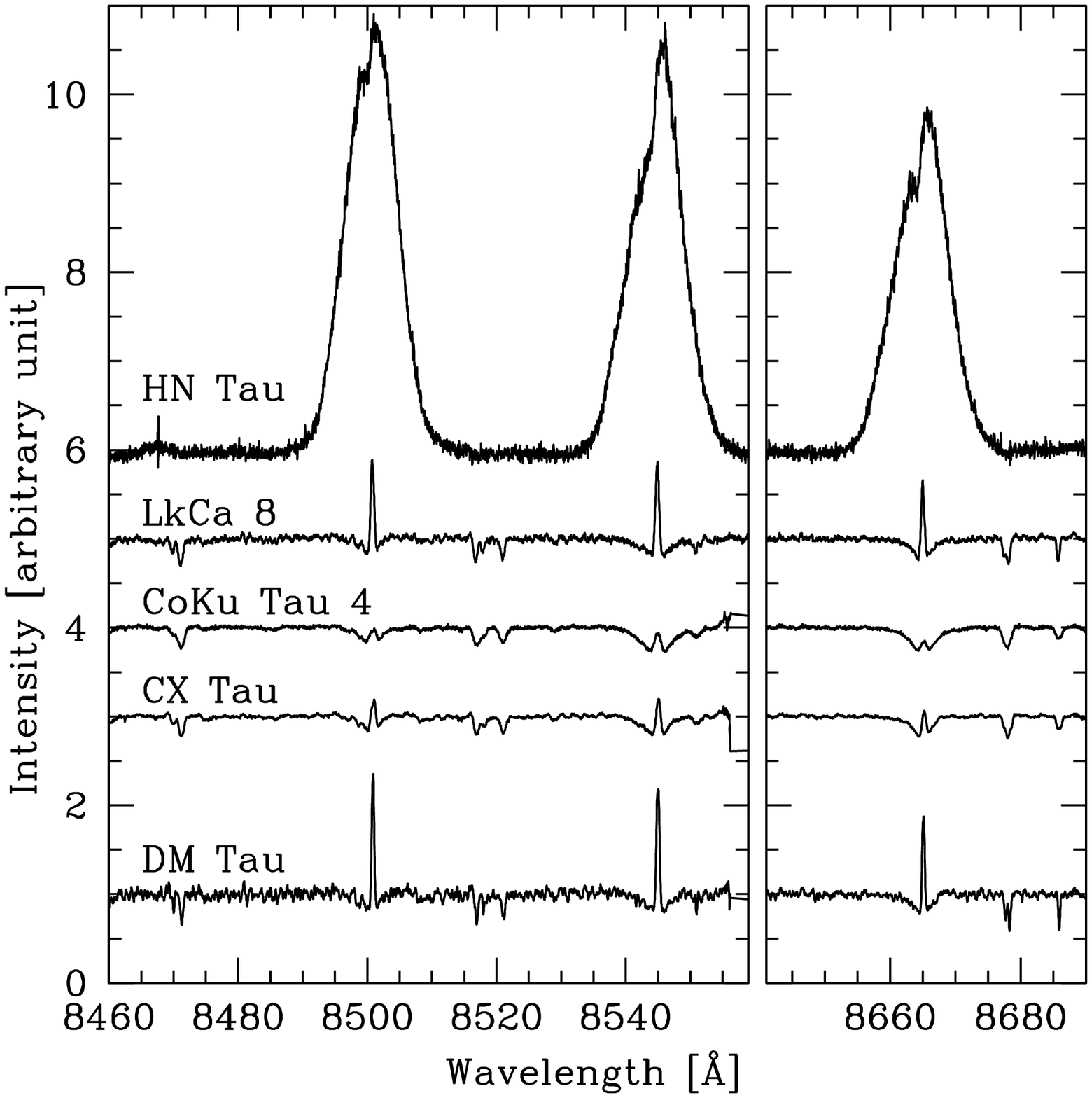}
 \caption{High resolution spectra of the YSOs.}
 \label{ap4}
\end{figure}

\begin{figure}[h]
 \centering
 \includegraphics[width=14.0cm, angle=0]{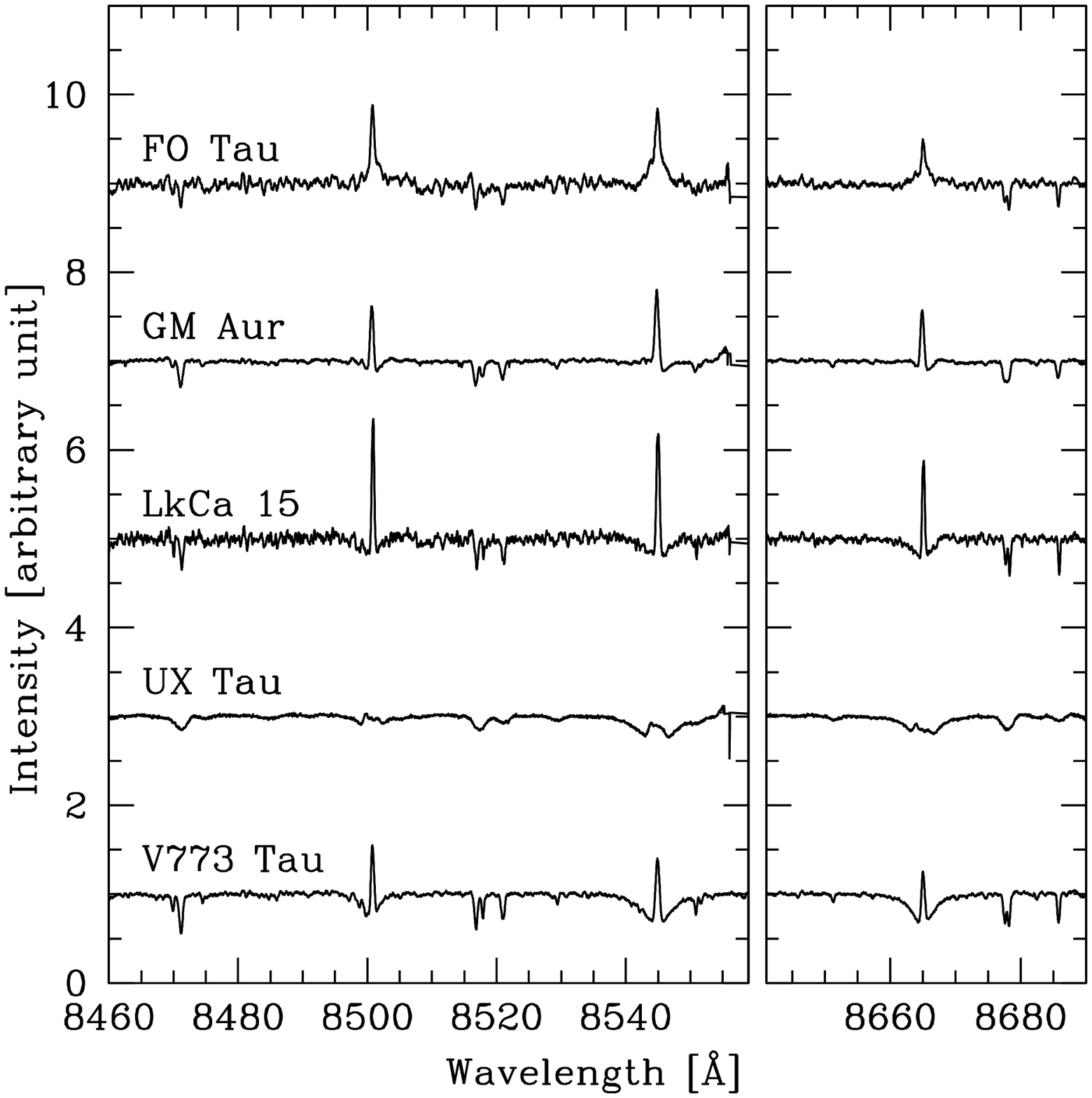}
 \caption{High resolution spectra of the YSOs.}
 \label{ap5}
\end{figure}

\begin{figure}[h]
 \centering
 \includegraphics[width=14.0cm, angle=0]{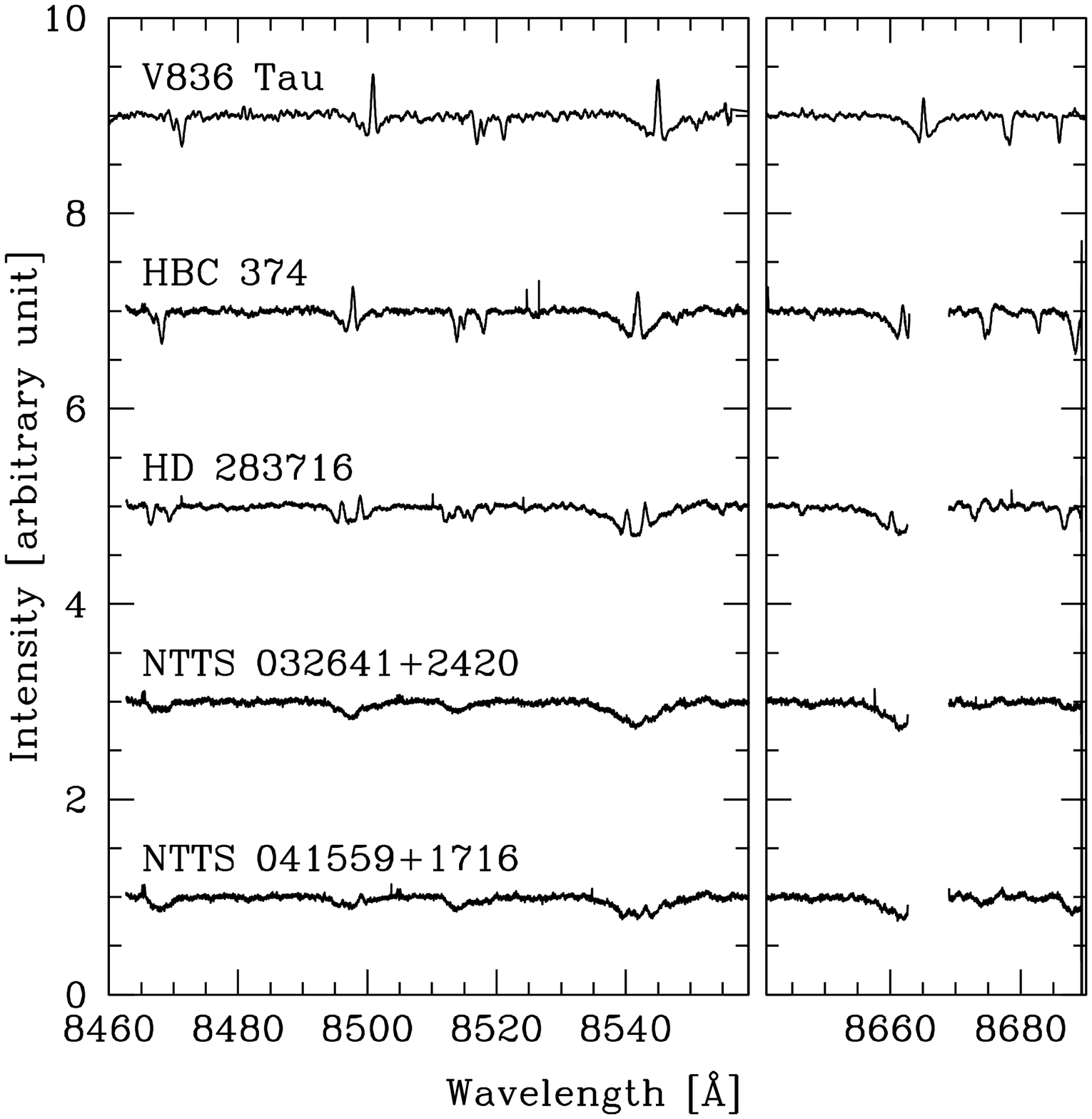}
 \caption{High resolution spectra of the YSOs.}
 \label{ap6}
\end{figure}

\begin{figure}[h]
 \centering
 \includegraphics[width=14.0cm, angle=0]{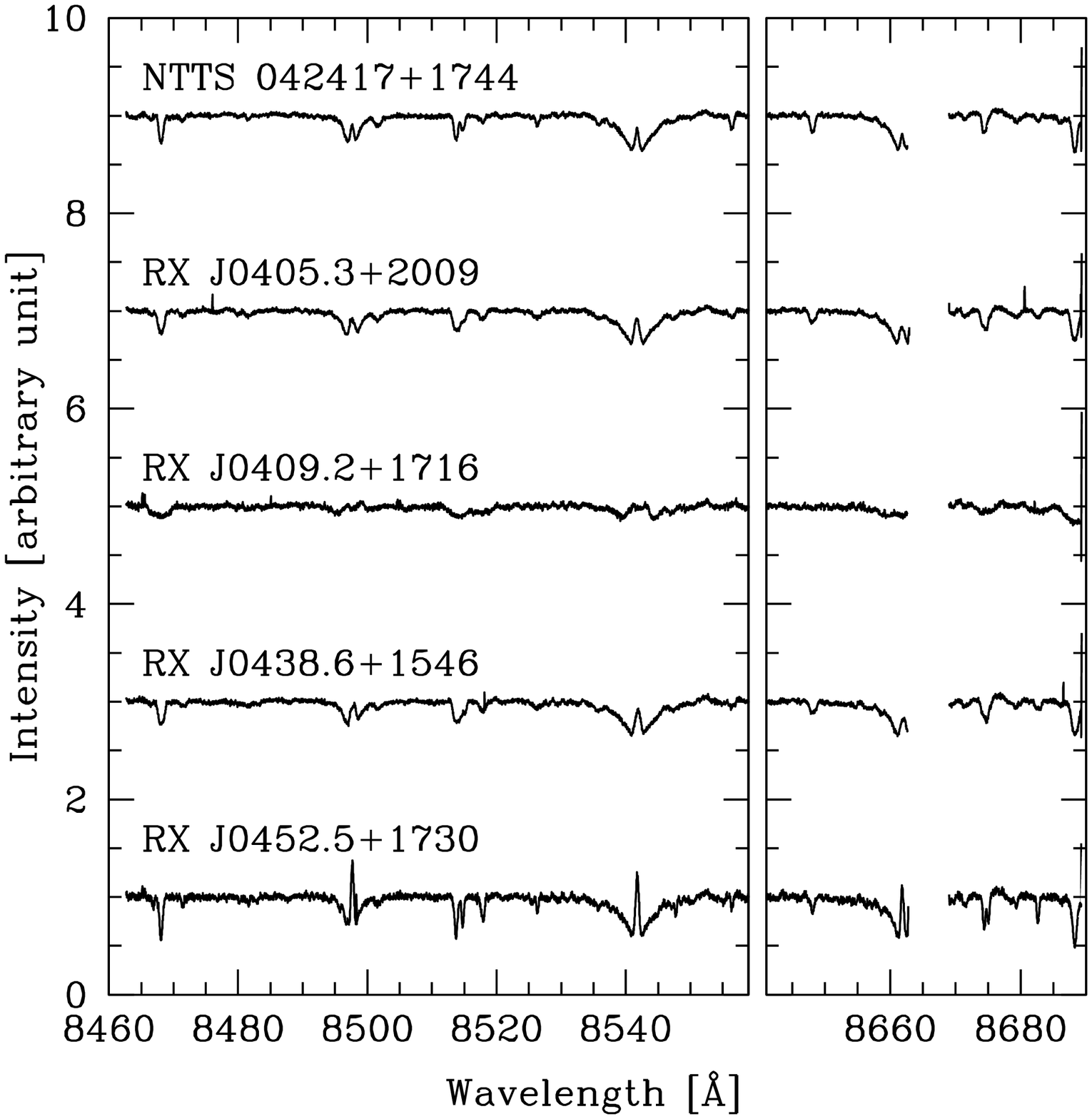}
 \caption{High resolution spectra of the YSOs.}
 \label{ap7}
\end{figure}

\begin{figure}[h]
 \centering
 \includegraphics[width=14.0cm, angle=0]{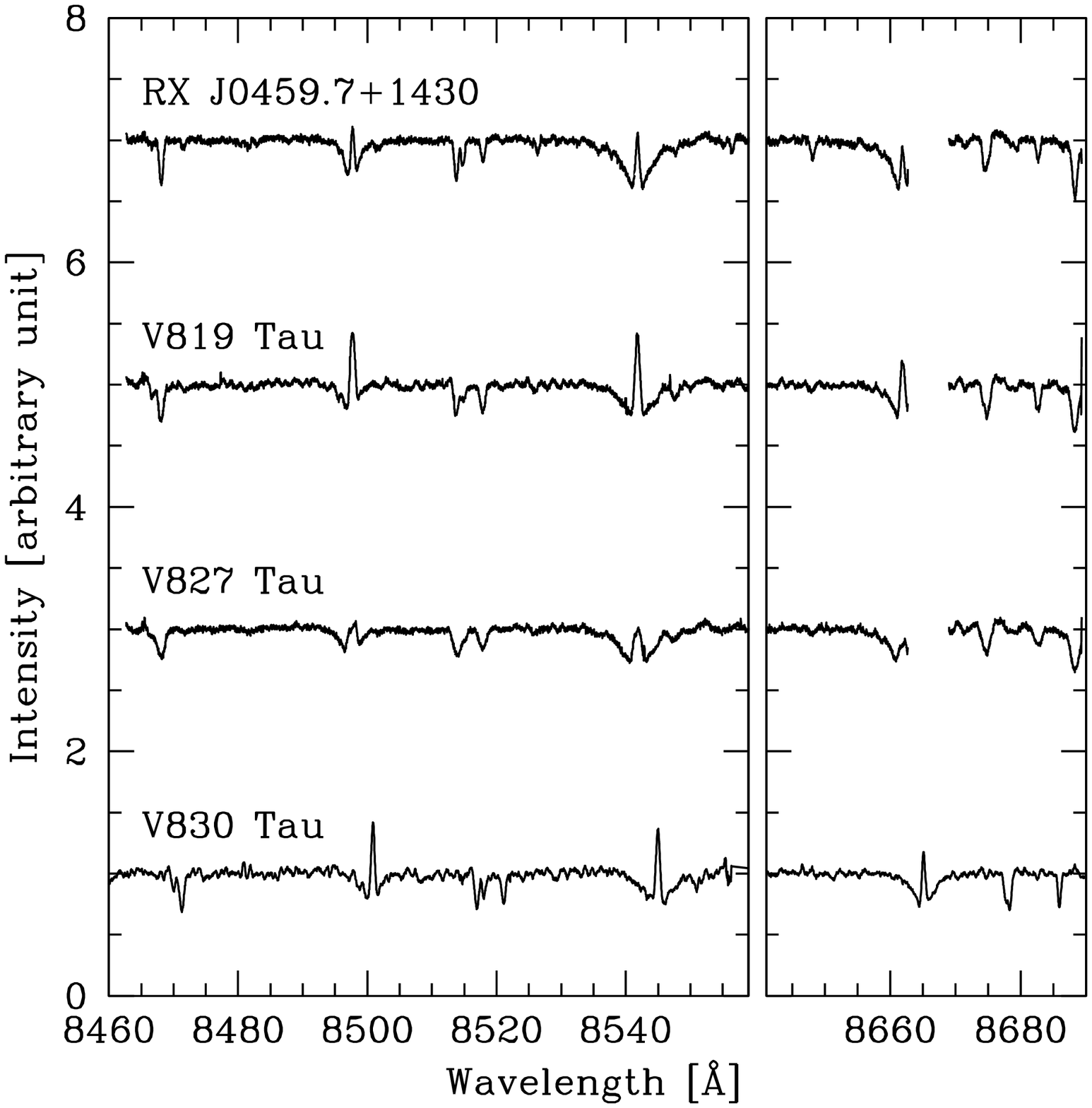}
 \caption{High resolution spectra of the YSOs.}
 \label{ap8}
\end{figure}

\clearpage

\bibliographystyle{raa}
\bibliography{ms1345}

\end{document}